\documentclass[a4paper]{jpconf}
\usepackage{graphicx,amsmath,dcolumn,bm}
\usepackage{wrapfig,epsfig}
\usepackage{epic,eepic}
\usepackage{feynmp}
\begin{document}

\title{Tensor-polarized structure functions: \\
Tensor structure of deuteron in 2020's}

\author{S. Kumano}

\address{KEK Theory Center,
             Institute of Particle and Nuclear Studies, KEK \\
             1-1, Ooho, Tsukuba, Ibaraki, 305-0801, Japan \\
             J-PARC Branch, KEK Theory Center,
             Institute of Particle and Nuclear Studies, KEK \\
           and
           Theory Group, Particle and Nuclear Physics Division, 
           J-PARC Center \\
           203-1, Shirakata, Tokai, Ibaraki, 319-1106, Japan}
           
\ead{shunzo.kumano@kek.jp}

\begin{abstract}
We explain spin structure for a spin-one hadron, in which 
there are new structure functions, in addition to the ones 
($F_1$, $F_2$, $g_1$, $g_2$) which exist for the spin-1/2 nucleon, 
associated with its tensor structure.
The new structure functions are $b_1$, $b_2$, $b_3$, and $b_4$
in deep inelastic scattering of a charged-lepton from
a spin-one hadron such as the deuteron. Among them, twist-two
functions are related by the Callan-Gross type relation
$b_2 = 2 x b_1$ in the Bjorken scaling limit.
First, these new structure functions are introduced, and 
useful formulae are derived for projection operators of $b_{1-4}$
from a hadron tensor $W_{\mu\nu}$. 
Second, a sum rule is explained for $b_1$, and
possible tensor-polarized distributions are discussed 
by using HERMES data in order to propose future experimental
measurements and to compare them with theoretical models. 
A proposal was approved to measure $b_1$ at 
the Thomas Jefferson National Accelerator Facility (JLab), 
so that much progress is expected for $b_1$ in the near future.
Third, formalisms of polarized proton-deuteron Drell-Yan processes
are explained for probing especially tensor-polarized antiquark
distributions, which were suggested by the HERMES data.
The studies of the tensor-polarized structure functions
will open a new era 
in 2020's for tensor-structure studies in terms of quark and gluon
degrees of freedom, which is much different from ordinary
descriptions in terms of nucleons and mesons.
\end{abstract}

\section{Introduction}

Spin structure of the nucleon has been extensively investigated for 
finding its origin. In a naive quark model, the spin-1/2 nucleon should
consist of two quarks with parallel spins to the nucleon spin and
a quark with antiparallel spin. 
In other words, quarks carry 100\% of the nucleon spin.
However, polarized lepton-nucleon scattering measurements clarified
that this simple description does not work.
The quarks carry only a small fraction of the nucleon spin. 
The remaining spin may be possibly carried by antiquark and 
gluon spins. However, antiquark spin contributions
are small and gluon spin contribution seems to be also small 
according to recent measurements and their global analyses.
Therefore, only remaining possibilities are orbital angular momentum
effects of quarks and gluons. They are currently investigated
by generalized parton distributions (GPDs) and 
transverse-momentum-dependent parton distributions (TMDs).
Therefore, understanding of orbital motion is crucial 
in hadron spin physics. 

For a spin-one hadron such as the deuteron, there exist new polarized 
structure functions, $b_1$, $b_2$, $b_3$, and $b_4$,
which could probe dynamical aspects including orbital motion
within the hadron. The new structure is associated with tensor structure,
so that the tensor-polarized structure functions vanish if constituents are 
in the $S$ state. It is also important to investigate completely
different spin quantities from the nucleon
for establishing high-energy spin physics.

\begin{figure}[t!]
   \vspace{-0.60cm}
   \begin{center}
   \includegraphics[width=0.65\textwidth]{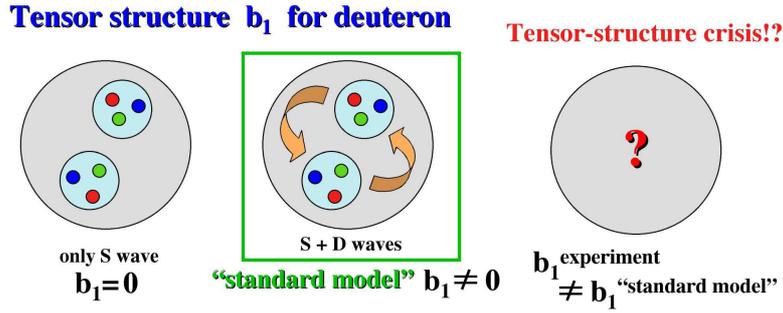}
   \end{center}
\vspace{-0.60cm}
\caption{Tensor structure of the deuteron}
\label{fig:b1-motivation}
\vspace{-0.3cm}
\end{figure}

The tensor structure of deuteron and other nuclei has been investigated
for a long time by nucleon and meson degrees of freedom
as illustrated in Fig.\,\ref{fig:b1-motivation}.
However, the tensor structure has not been explored yet
in terms of quark and gluon degrees 
of freedom. There are some studies. First, the new structure functions
$b_{1-4}$ are introduced for a spin-one hadron \cite{fs83,hjm89}.
For the leading-twist structure function $b_1$, a useful sum rule
was proposed by using a parton model \cite{b1-sum}.
Leptoproduction of spin-one hadrons is investigated in 
Ref. \cite{rho-production}.
There are theoretical studies on the $x$ distribution of $b_1$.
If the constituents are in the S wave, it vanishes ($b_1=0$)
\cite{hjm89}.
A finite distribution with a node is expected in a convolution model
with D-state admixture \cite{hjm89,kh91},
effects of pions and six-quark configuration 
\cite{miller-b1}, and shadowing effects in a nucleus
\cite{b1-shadowing,bj98}.
If we find a significant difference from these theoretical
expectations, the studies could create a new field of
high-energy spin physics.
There are related theoretical studies such as new fragmentation 
functions \cite{spin-1-frag}, generalized parton distributions 
\cite{spin-1-gpd}, target mass corrections \cite{mass-corr},
positivity constraints \cite{dmitrasinovic-96},
lattice QCD estimate \cite{lattice},
projection operators of $b_{1-4}$ \cite{kk08} for spin-one hadrons,
and angular momenta for spin-1 hadron \cite{angular-spin-1}.
As an alternative reaction for investigating the tensor structure
is to use Drell-Yan processes.
A theoretical formalism was developed in 
Ref. \cite{pd-drell-yan} to study the tensor-polarized
distributions at hadron facilities by Drell-Yan processes
with polarized deuteron.

The HERMES collaboration reported the measurement of the structure 
function $b_1$ in 2005 \cite{hermes05}. The data indicated 
a finite distribution at $x<0.1$, which roughly agrees with
a double scattering contribution estimated in Ref. \cite{bj98}.
The data are also consistent with the quark-parton model sum rule
for $b_1$ \cite{b1-sum} although experimental errors are still large.
As a future experiment, the proposal was approved for measuring $b_1$
at Thomas Jefferson National Accelerator Facility (JLab) \cite{Jlab-b1}.
It could be also investigated at the Electron-Ion Collider (EIC)
\cite{eic}. Tensor polarization $A_{zz}$ at large-$x$ ($x>1$)
can be also investigated \cite{azz}. In addition,
the tensor-polarized quark and antiquark distributions could be 
studied by Drell-Yan processes with tensor-polarized deuteron
at Fermilab \cite{fermilab-d}, 
J-PARC (Japan Proton Accelerator Research Complex) \cite{j-parc},
and GSI-FAIR (Gesellschaft f\"ur Schwerionenforschung -Facility
for Antiproton and Ion Research) \cite{gsi-fair}.

In this article, we explain the definition of tensor-polarized
structure functions in Sec.\,\ref{b1-4-definition}.
Then, operators to project out structure functions of a spin-1
hadron are shown in Sec.\,\ref{projection}. A sum rule of $b_1 (x)$
is explained in Sec.\,\ref{sum_rule}. A useful parametrization 
is proposed for tensor-polarized distributions in 
Sec.\,\ref{parametrization} for explaining the HERMES data.
As shown in Sec.\,\ref{pd-drell-yan},
the tensor-polarized distributions could be also investigated by
Drell-Yan processes with tensor-polarized deuteron \cite{pd-drell-yan}.
The results are summarized  in Sec.\,\ref{summary}.


\section{Tensor polarized structure functions}
\label{b1-4-definition}

\vspace{0.2cm}

\begin{wrapfigure}[12]{r}{0.40\textwidth}
   \vspace{-0.7cm}
   \begin{center}
       \epsfig{file=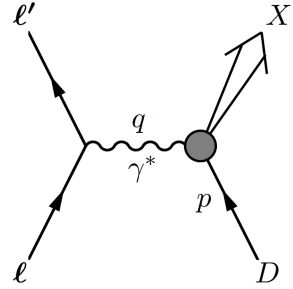,width=3.8cm}
   \end{center}
   \vspace{-0.5cm}
   \caption{Deep inelastic scattering of charged lepton from a hadron $D$.}
\label{fig:electron-nucleon-dis}
\end{wrapfigure}

We consider charged-lepton deep inelastic scattering (DIS) 
from a hadron in Fig.\,\ref{fig:electron-nucleon-dis}.
Its cross section is described by a hadron 
tensor $W_{\mu \nu}$ multiplied by a lepton tensor $L^{\mu\nu}$.
The hadron tensor for the spin-1/2 nucleon is 
described by four structure functions $F_1$, $F_2$, $g_1$,
and $g_2$:
\begin{align}
W_{\mu\nu}^{\lambda_f \lambda_i} = -F_1 & \hat{g}_{\mu \nu} 
+\frac{F_2}{M \nu} \hat{p}_\mu \hat{p}_\nu 
+ \frac{i g_1}{\nu}\epsilon_{\mu \nu \lambda \sigma} q^\lambda s^\sigma  
+\frac{i g_2}{M \nu ^2}\epsilon_{\mu \nu \lambda \sigma} 
       q^\lambda (p \cdot q s^\sigma - s \cdot q p^\sigma ),
\\
& 
\hat{g}_{\mu \nu} \equiv  g_{\mu \nu} -\frac{q_\mu q_\nu}{q^2}, \ \ 
\hat{a}_\mu \equiv a_\mu -\frac{a \cdot q}{q^2} q_\mu ,
\label{eqn:w-1/2}
\end{align}

\noindent
where $\epsilon_{\mu \nu \lambda \sigma}$ is an antisymmetric tensor
with the convention $\epsilon_{0123}=+1$, $\nu$ is defined by
$\nu ={p \cdot q}/{M}$ with the hadron mass $M$, hadron momentum $p$,
and momentum transfer $q$, $Q^2$ is given by $Q^2=-q^2>0$,
and $s^\mu$ is the spin vector with the constraint $s \cdot p=0$.
The notations $\hat{g}_{\mu \nu}$ and $\hat{p}_\mu$ are introduced
so as to ensure the current conservation 
$q^\mu W _{\mu \nu} = q^\nu W _{\mu \nu}=0$. 
The initial and final spin states are denoted by $\lambda_i$
and $\lambda_f$, respectively. In general, off-diagonal terms 
with $\lambda_f \ne \lambda_i$ are needed to consider
higher-twist contributions \cite{hjm89}.

In a spin-one hadron, there are four additional structure functions
$b_1$, $b_2$, $b_3$ and $b_4$ in the hadron tensor \cite{hjm89,kk08}:
\begin{align}
W_{\mu \nu}^{\lambda_f \lambda_i}
   = & -F_1 \hat{g}_{\mu \nu} 
     +\frac{F_2}{M \nu} \hat{p}_\mu \hat{p}_\nu 
     + \frac{ig_1}{\nu}\epsilon_{\mu \nu \lambda \sigma} q^\lambda s^\sigma  
     +\frac{i g_2}{M \nu ^2}\epsilon_{\mu \nu \lambda \sigma} 
      q^\lambda (p \cdot q s^\sigma - s \cdot q p^\sigma )
\notag \\
& 
     -b_1 r_{\mu \nu} 
     + \frac{1}{6} b_2 (s_{\mu \nu} +t_{\mu \nu} +u_{\mu \nu}) 
     + \frac{1}{2} b_3 (s_{\mu \nu} -u_{\mu \nu}) 
     + \frac{1}{2} b_4 (s_{\mu \nu} -t_{\mu \nu}) ,
\label{eqn:w-1}
\end{align}
where $r_{\mu \nu}$, $s_{\mu \nu}$, $t_{\mu \nu}$, and $u_{\mu \nu}$
are defined by
\begin{align}
r_{\mu \nu} = & \frac{1}{\nu ^2}
   \bigg [ q \cdot E ^* (\lambda_f) q \cdot E (\lambda_i) 
           - \frac{1}{3} \nu ^2  \kappa \bigg ]
   \hat{g}_{\mu \nu}, 
\ \ \ \ 
s_{\mu \nu} =  \frac{2}{\nu ^2} 
   \bigg [ q \cdot E ^* (\lambda_f) q \cdot E (\lambda_i) 
           - \frac{1}{3} \nu ^2  \kappa \bigg ]
\frac{\hat{p}_\mu \hat{p}_\nu}{M \nu}, \notag \\
t_{\mu \nu} = & \frac{1}{2 \nu ^2}
   \bigg [ q \cdot E ^* (\lambda_f) 
           \left\{ \hat{p}_\mu \hat E_\nu (\lambda_i) 
                 + \hat{p} _\nu \hat E_\mu (\lambda_i) \right\}
   + \left\{ \hat{p}_\mu \hat E_\nu^* (\lambda_f)  
           + \hat{p}_\nu \hat E_\mu^* (\lambda_f) \right\}  
     q \cdot E (\lambda_i) 
   - \frac{4 \nu}{3 M}  \hat{p}_\mu \hat{p}_\nu \bigg ] ,
\notag \\
u_{\mu \nu} = & \frac{M}{\nu} 
   \bigg [ \hat E_\mu^* (\lambda_f) \hat E_\nu (\lambda_i) 
          +\hat E_\nu^* (\lambda_f) \hat E_\mu (\lambda_i) 
   +\frac{2}{3}  \hat{g}_{\mu \nu}
   -\frac{2}{3 M^2} \hat{p}_\mu \hat{p}_\nu \bigg ] .
\end{align}
Here, $\kappa$ is given by $\kappa= 1+{Q^2}/{\nu^2}$, 
and $s^\mu$ is the spin vector of the spin-one hadron.
The $E^\mu$ is the polarization vector of the spin-one hadron 
and it satisfies the conditions, $p \cdot E =0$ and $E^* \cdot E =-1$.
It is taken as the spherical unit vectors:
\begin{align}
  E^\mu (\lambda= \pm 1) = \frac{1}{\sqrt{2}}(0,\mp 1, -i,0),
\ \ \ 
  E^\mu (\lambda=0) = (0,0,0,1) .
\end{align}
Then, the spin vector $s$ is given by the polarization vector as
\begin{equation}   
(s_{\lambda_f \lambda_i})^{\mu}
      = -\frac{i}{M} \epsilon ^{\mu \nu \alpha \beta} 
                E^*_\nu (\lambda_f) E_\alpha (\lambda_i) p_\beta ,
\end{equation}
where $M$ is the mass of the spin-1 hadron.
The initial and final polarization vectors are denoted by
$E^\mu (\lambda_i)$ and $E^\mu (\lambda_f)$, respectively,
with the spin states $\lambda_i$ and $\lambda_f$.
Explicit expressions for the spin vector $s^\mu$ are given
for $s^\mu _{1 1}$, $s^\mu _{1 0}$, and $s^\mu _{0 1}$ as
\begin{equation}
s^\mu _{1 1}   = (0,0,0,1), \ \  
s^\mu _{1 0}   = \frac{1}{\sqrt{2}} (0,1,-i,0), \ \ 
s^\mu _{0 1}   = (s^\mu _{1 0})^* .
\end{equation}

In Eq. (\ref{eqn:w-1}),
the coefficients of $b_1$, $b_2$, $b_3$ and $b_4$ are symmetric under 
$\mu \leftrightarrow \nu$, and they vanish under the spin average. 
The coefficients are defined so that $b_1$ and $b_2$ are twist-two
functions and they satisfy the Callan-Gross type relation.
In addition to the functions $F_1$, $F_2$, $g_1$, and $g_2$, which
exist in a spin-1/2 hadron, 
there are four new structure functions, $b_1$, $b_2$, $b_3$ and $b_4$.
They are related to the tensor structure of the spin-1/2 hadron.

\section{Projection operators for tensor-polarized structure functions}
\label{projection}

\vspace{0.2cm}

In a convolution description of nuclear structure functions,
the hadron tensor $W_{\mu\nu}$ is calculated by the convolution integral
of nucleonic $W_{\mu\nu}$ with a lightcone momentum distribution of
a nucleon in a nucleus. Then, the nuclear structure function $F_2^A$
is, for example, extracted from the nuclear tensor $W_{\mu\nu}^A$ 
by applying a projection operator to $W_{\mu\nu}^A$.
Because there are eight structure functions for a spin-one hadron,
eight independent combinations are needed for tensors with
indices $\mu$ and $\nu$. Only $\lambda=1$ and $\lambda=0$ terms 
are used because $\lambda=-1$ terms make the same contributions 
as $\lambda=1$ ones. 
We choose the following terms
\vspace{-0.3cm}
\begin{gather}
  g^{\mu\nu} \delta_{\lambda_f  1}
             \delta_{\lambda_i  1}   , \ \ 
  g^{\mu\nu} \delta_{\lambda_f  0}
             \delta_{\lambda_i  0}   , \ \ 
  g^{\mu\nu} \delta_{\lambda_f  1}
             \delta_{\lambda_i  0}   , \ \ 
  \frac{p^\mu p^\nu}{M^2} \delta_{\lambda_f  1}
                          \delta_{\lambda_i  1}   , \ \ 
  \frac{p^\mu p^\nu}{M^2} \delta_{\lambda_f  0}
                          \delta_{\lambda_i  0}   ,
\notag \\
  \frac{1}{M}
  [ p^\mu E ^\nu (\lambda =1) 
   +p^\nu E ^\mu (\lambda =1) ] \,  
   \delta_{\lambda_f  1}  \delta_{\lambda_i  0}   , \ \ 
  \frac{i}{M}
  \epsilon ^{\mu \nu \alpha \beta } 
  q_\alpha s_\beta ^{11}  \delta_{\lambda_f  1}
                          \delta_{\lambda_i  1} , \ \ 
  \frac{i}{M}
  \epsilon ^{\mu \nu \alpha \beta } 
  q_\alpha s_\beta ^{10}  \delta_{\lambda_f  0}
                          \delta_{\lambda_i  1} .
\end{gather}
Then, we obtain the projection operators for the structure functions
of a spin-one hadron as \cite{kk08}
\begin{align}
F_1 & = - \frac{1}{2} 
          \bigg ( g^{\mu \nu} 
                  - \frac{\kappa-1}{\kappa} \frac{p^\mu p^\nu}{M^2} \bigg )
   \frac{1}{3} 
   \delta_{\lambda_f \lambda_i}
   W_{\mu \nu} ^{\lambda_f \lambda_i} ,
\notag \\    
F_2 & = -\frac{x}{\kappa} 
        \bigg (g ^{\mu\nu} 
               - \frac{\kappa-1}{\kappa} \frac{3 p^\mu p^\nu}{M^2} \bigg ) 
   \frac{1}{3} 
   \delta_{\lambda_f \lambda_i}
   W_{\mu \nu} ^{\lambda_f \lambda_i} , 
\notag \\
g_1 & = - \frac{i}{2 \kappa \nu} \epsilon ^{\mu \nu \alpha \beta} q_\alpha
         \bigg (  s_\beta ^{11} \delta_{\lambda_f  1}
                                \delta_{\lambda_i  1} 
                - s_\beta ^{10} \delta_{\lambda_f  0}
                                \delta_{\lambda_i  1} 
         \bigg ) W_{\mu \nu} ^{\lambda_f \lambda_i} , 
\notag \\
g_2 & =  \frac{i}{2 \kappa \nu} \epsilon ^{\mu \nu \alpha \beta} q_\alpha
         \bigg (  s_\beta ^{11}  \delta_{\lambda_f  1}
                                 \delta_{\lambda_i  1} 
                + \frac{s_\beta ^{10}}{\kappa-1}
                                \delta_{\lambda_f  0}
                                \delta_{\lambda_i  1} 
         \bigg ) W_{\mu \nu} ^{\lambda_f \lambda_i} , \ \ \ 
\notag \\
b_1 & = \bigg [ -\frac{1}{2 \kappa}  g^{\mu \nu}
        \big ( \delta_{\lambda_f  0}
               \delta_{\lambda_i  0} 
              -\delta_{\lambda_f  1}
               \delta_{\lambda_i  0}  \big ) 
  + \frac{\kappa-1}{2 \kappa^2}  \frac{p^\mu p^\nu}{M^2} 
        \big ( \delta_{\lambda_f  0}
               \delta_{\lambda_i  0} 
              -\delta_{\lambda_f  1}
               \delta_{\lambda_i  1}  \big ) 
        \bigg ] W_{\mu \nu} ^{\lambda_f \lambda_i} , 
\notag \\
b_2 & = \frac{x}{\kappa^2} \bigg [
        g^{\mu\nu} \bigg \{ - \delta_{\lambda_f  0}
                              \delta_{\lambda_i  0} 
                            - 2 (\kappa-1) \delta_{\lambda_f  1}
                                           \delta_{\lambda_i  1} 
                            + (2\kappa-1)  \delta_{\lambda_f  1}
                                           \delta_{\lambda_i  0} 
                   \bigg \}
\notag \\
  &  \! \! \! \! \!  
         +\frac{3(\kappa-1)}{\kappa}  \frac{p^\mu p^\nu}{M^2} 
                 \big ( \delta_{\lambda_f  0}
                        \delta_{\lambda_i  0} 
                       -\delta_{\lambda_f  1}
                        \delta_{\lambda_i  1}  \big )
       -\frac{4 (\kappa-1)}{\sqrt{\kappa} M}
          \big \{ p^\mu E^\nu (\lambda =1) + p^\nu E^\mu (\lambda =1) \big \}
                        \delta_{\lambda_f  1}
                        \delta_{\lambda_i  0}  \bigg ] 
        W_{\mu \nu} ^{\lambda_f \lambda_i},
\notag \\
b_3 & = \frac{x}{3 \kappa^2} \bigg [
        g^{\mu\nu} \bigg \{ - \delta_{\lambda_f  0}
                              \delta_{\lambda_i  0} 
                            + \frac{2(2\kappa^2+2\kappa-1)}{\kappa-1} 
                              \delta_{\lambda_f  1}
                              \delta_{\lambda_i  1} 
                            - \frac{4\kappa^2+3\kappa-1}{\kappa-1}
                              \delta_{\lambda_f  1}
                              \delta_{\lambda_i  0} 
                   \bigg \}
\notag \\
  &  \! \! \! \! \!  
       +\frac{3(\kappa-1)}{\kappa}  \frac{p^\mu p^\nu}{M^2} 
                 \big (  \delta_{\lambda_f  0}
                         \delta_{\lambda_i  0} 
                       - \delta_{\lambda_f  1}
                         \delta_{\lambda_i  1}  \big )
       -\frac{4 (\kappa-1)}{\sqrt{\kappa} M}
          \big \{ p^\mu E^\nu (\lambda =1) + p^\nu E^\mu (\lambda =1) \big \}
                              \delta_{\lambda_f  1}
                              \delta_{\lambda_i  0}  \bigg ] 
        W_{\mu \nu} ^{\lambda_f \lambda_i},
\notag \\
b_4 & = \frac{x}{3 \kappa^2} \bigg [
        g^{\mu\nu} \bigg \{ - \delta_{\lambda_f  0}
                              \delta_{\lambda_i  0} 
                            - \frac{2(\kappa^2+4\kappa+1)}{\kappa-1} 
                              \delta_{\lambda_f  1}
                              \delta_{\lambda_i  1} 
                            + \frac{2\kappa^2+9\kappa+1}{\kappa-1} 
                              \delta_{\lambda_f  1}
                              \delta_{\lambda_i  0} 
                   \bigg \}
\notag \\
  &  \! \! \! \! \!  
       +\frac{3(\kappa-1)}{\kappa}  \frac{p^\mu p^\nu}{M^2} 
                 \big (  \delta_{\lambda_f  0}
                         \delta_{\lambda_i  0} 
                       - \delta_{\lambda_f  1}
                         \delta_{\lambda_i  1}  \big )  
       +\frac{4(2\kappa+1)}{\sqrt{\kappa} M}
          \big \{ p^\mu E^\nu (\lambda =1) + p^\nu E^\mu (\lambda =1) \big \}
                              \delta_{\lambda_f  1}
                              \delta_{\lambda_i  0}  \bigg ] 
        W_{\mu \nu} ^{\lambda_f \lambda_i} ,
\label{eqn:spin-1-proj}
\end{align}
where summations are taken over $\lambda_i$ and $\lambda_f$.

Because leading-twist structure functions $b_1$ and $b_2$ are
experimentally measured first, it is useful to consider 
the Bjorken scaling limit, $\nu, \, Q^2 \to \infty$ with 
finite $x=Q^2/(2 p \cdot q)$. In this limit, we have the relations
\begin{align}
&   \lim_{\text{Bj}} g^{\mu\nu} W_{\mu \nu} ^{10} 
 = \lim_{\text{Bj}} g^{\mu\nu} W_{\mu \nu} ^{11} ,
\ \ \ \ 
   \lim_{\text{Bj}} \, (\kappa-1)
    \frac{p^\mu p^\nu}{M^2} W_{\mu \nu} ^{\lambda \lambda} = 0 ,
\nonumber \\
&  \lim_{\text{Bj}} \,
    \frac{\kappa-1}{M} [ \, p^\mu E^\nu (\lambda=1)
                 +p^\nu E^\mu (\lambda=1) \, ] \, W_{\mu \nu} ^{10} = 0 ,
\end{align}
by noting $\kappa \rightarrow 1$, $2x F_1 \rightarrow F_2$, and.
$2x b_1 \rightarrow b_2$. Then, we obtain the expressions
from the projection operator in the Bjorken scaling limit as
\begin{align}
F_1 & =   \frac{1}{2x} F_2 =  
   - \frac{1}{2} g^{\mu \nu} 
     \frac{1}{3} 
     \delta_{\lambda_f \lambda_i}
     W_{\mu \nu} ^{\lambda_f \lambda_i} , \ \ \ 
g_1 = 
  - \frac{i}{2 \nu}
  \epsilon ^{\mu \nu \alpha \beta} 
  q_\alpha s_\beta ^{11} 
  \delta_{\lambda_f 1} \delta_{\lambda_i 1} 
  W_{\mu \nu} ^{\lambda_f \lambda_i} ,
\notag \\
b_1 & = \frac{1}{2x} b_2 =  
   \frac{1}{2} g^{\mu \nu}
   \big ( \delta_{\lambda_f 1} \delta_{\lambda_i 1} 
         -\delta_{\lambda_f 0} \delta_{\lambda_i 0}  \big ) 
   W_{\mu \nu} ^{\lambda_f \lambda_i} .
\label{eqn:bj_scaling-pro}
\end{align}
The exact relations of Eq. (\ref{eqn:spin-1-proj}) or
the ones of Eq. (\ref{eqn:bj_scaling-pro}) in the scaling limit
are useful equations for extracting the new structure functions
$b_{1-4}$ in addition to the usual ones, $F_{1,2}$ and $g_{1,2}$,
for estimating them in theoretical descriptions, especially in
the convolution model.

\section{Sum rule for $b_1$}
\label{sum_rule}

\vspace{0.2cm}

\begin{wrapfigure}[10]{r}{0.48\textwidth}
   \vspace{-0.8cm}
   \begin{center}
       \epsfig{file=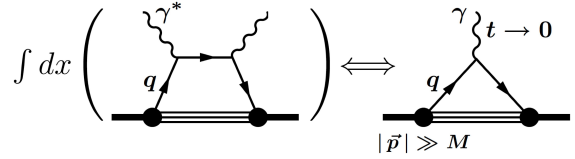,width=7.5cm}
   \end{center}
   \vspace{-0.7cm}
\caption{Structure function and elastic form factor 
by the parton model in the frame with $|\vec p \,| \gg M$.}
\label{fig:b1sum}
\end{wrapfigure}

We discuss a useful sum rule for the structure function $b_1$.
Sum rules of structure functions could be derived in
a parton model by considering the relation between 
the structure function integrated over $x$ and elastic form factor
in the infinite momentum frame \cite{feynman}.
In the parton model, the structure function $b_1$ is expressed
by the tensor-polarized distributions $\delta_{_T} q (x)$ as
\cite{hjm89,b1-sum,factor-2}
\begin{align}
b_1 (x,Q^2) = \frac{1}{2} \sum_i e_i^2 
      \, \left [ \delta_{_T} q_i (x,Q^2) 
      + \delta_{_T} \bar q_i (x,Q^2)   \right ] , \ \ \
       \delta_{_T} q_i  \equiv q_i^0 - \frac{q_i^{+1}+q_i^{-1}}{2} ,
\label{eqn:b1-parton}
\end{align}
where $i$ indicates the flavor of a quark and $e_i$ is its charge,
and $q_i^\lambda$ indicates an unpolarized-quark distribution
in the hadron spin state $\lambda$. 
Instead of $\delta_{_T}$, $\delta$ could be used for the tensor-polarized
distributions. Because $\delta$ and $\Delta_{_T}$ are often used
for transversity distributions of spin-1/2 nucleon, the notation
$\delta_{_T}$ is used throughout this article.
The function $\delta_{_T} q$ indicates
an unpolarized-quark distribution in a tensor-polarized
spin-one hadron. The idea how to derive a sum rule for $b_1$ 
is illustrated in Fig.\,\ref{fig:b1sum}. 
The hadron tensor is equal to the imaginary part of
the forward scattering amplitude for virtual photon scattering
from the hadron by the optical theorem.
An integral of a DIS structure function is related to 
the form factor in the infinite momentum frame by
expressing the form factor in terms of parton momentum 
distributions as discussed in Ref. \cite{feynman}, where
the Bjorken sum rule is shown by this method.

Before stepping into the details, we may estimate the sum rule 
of $b_1 (x)$ intuitively by using the dimensional counting. 
The $b_1$ is defined in Eq. (\ref{eqn:w-1}) as a dimensionless
quantity. If it is integrated over $x$, it should become 
electromagnetic quantities of the deuteron probed by the charged lepton.
Due to the parity conservation, the new quantity, which does
not exist in the spin-1/2 nucleon, should be the electric quadrupole
moment. The electric quadrupole moment has the dimension of 
length$^2$=1/mass$^2$ except for the charge factor. 
Therefore, the $b_1$ sum is expected to be 
$\int dx b_1 (x) \sim \text{(mass dim.)}^2 \cdot Q_h$, where $Q_h$ is 
the electric quadrupole moment of the hadron $h$.
The quantity of the mass dimension may be the deuteron mass $M$,
but it is not obvious, so we 
proceed to a more detailed calculation.

For the deuteron, we define the structure function $b_1$ and 
parton distribution functions (PDFs) by the ones per nucleon
($b_1 /2 \to b_1$, $q (x)/2 \to q(x)$)
Taking the integral of Eq. (\ref{eqn:b1-parton}) over $x$ 
and then using the relations, 
$ ( \delta_{_T} u_v )_D \equiv ( \delta_{_T} u - \delta_{_T} \bar u )_D
                  = (\delta_{_T} u_v^p + \delta_{_T} u_v^n)/2
                  = (\delta_{_T} u_v + \delta_{_T} d_v)/2$,
$ ( \delta_{_T} d_v )_D = (\delta_{_T} d_v + \delta_{_T} u_v)/2$,             
for valence quark distributions, we obtain
\begin{align}
\int dx \, b_1 (x) & 
      = \frac{5}{36} \int dx
      \, \left [ \, \delta_{_T} u_v (x) + \delta_{_T} d_v (x) \, \right ]
\nonumber \\
    & + \frac{1}{18} \int dx
      \, \left [ \, 8 \, \delta_{_T} \bar u (x) + 2\,  \delta_{_T} \bar d (x) 
                    + \delta_{_T}      s (x) +   \delta_{_T} \bar s (x) 
      \, \right ]_D , 
\label{eqn:b1-sum-parton-1}
\end{align}
where $Q^2$ dependence is abbreviated.

For describing the elastic scattering, we define the helicity amplitude
with the charge operator $J_0$ by 
\begin{align}
\Gamma_{H,H} = \langle \, p, \, H \, \left | \, J_0 (0) 
                     \, \right | \, p, \, H \, \rangle .
\label{eqn:hecility-amp-1}
\end{align}
As shown in Fig.\,\ref{fig:b1sum}, the frame with large longitudinal
momentum $ | \, \vec p \, | \gg M$ is considered. Then,
the paton model can be used for estimating the amplitude
by assuming that the quarks move along the longitudinal direction
with the momentum fraction $x$, and we obtain
\begin{align}
\Gamma_{H,H} = \sum_i e_i \int dx
\, \left [ \, q_i^H (x) - \bar q_i ^H (x) \, \right ]_D
\equiv \sum_i e_i \int dx \, 
\left [ \, q_{i,v}^H (x) \, \right ]_D .
\label{eqn:hecility-amp-1}
\end{align}
The tensor spin combination of the amplitudes is written
in terms of the valence-quark distributions in the deuteron 
by the proton and neutron contributions
\begin{align}
\Gamma_{0,0} - \frac{\Gamma_{1,1} + \Gamma_{-1,-1} }{2}
= \frac{1}{3} \int dx \, \left [ \, \delta_{_T} u_v (x) 
                                  + \delta_{_T} d_v (x)   \right ] ,
\label{eqn:hecility-amp-2}
\end{align}
where the relation
$\int dx \left [ \delta_{_T} s (x) - \delta_{_T} \bar s (x) \right ] = 0$
is used. Substituting this relation into Eq. (\ref{eqn:b1-sum-parton-1}),
we obtain
\begin{align}
\int dx \, b_1 (x) 
      = \frac{5}{12} 
\, \left [ \, \Gamma_{0,0} - \frac{ \Gamma_{1,1} + \Gamma_{-1,-1} }{2} \, \right ]
      + \frac{1}{9} \int dx
      \, \left [ \, 4 \, \delta_{_T} \bar u (x) +   \delta_{_T} \bar d (x) 
                     +   \delta_{_T} \bar s (x)  \, \right ]_D .
\label{eqn:b1-sum-parton-2}
\end{align}

The next step is to calculate the tensor helicity combination
by macroscopic quantities for the deuteron.
The elastic scattering amplitudes can be described 
in terms of electric charge and quadrupole
form factors, $F_C (t)$ and $F_Q (t)$ where $t$ is 
the momentum-transfer squared, of the deuteron as
\begin{align}
\Gamma_{0,0} = 
\lim_{t \to 0} 
\left [ F_C (t) - \frac{t}{3 M^2} F_Q (t)  \right ] , \ \ \ 
\Gamma_{1,1} = \Gamma_{-1,-1} = 
\lim_{t \to 0} 
     \left [ F_C (t) + \frac{t}{6 M^2} F_Q (t)  \right ] .
\label{eqn:form-factor-1}
\end{align}
Here, $t \to 0$ limit is taken, and the units of $F_C$ and $F_Q$
are given by the unit charge $e$ and the mass of the deuteron $M$
as $e$ and $e/M^2$, respectively. 
Then, the tensor spin combination is expressed 
in terms of the quadrupole form factor as
\begin{align}
\Gamma_{0,0} - \frac{\Gamma_{1,1} + \Gamma_{-1,-1} }{2}
= - \lim_{t \to 0} \frac{t}{4} F_Q (t)
= 0 .
\label{eqn:hecility-amp-3}
\end{align}
Substituting this relation into Eq. (\ref{eqn:b1-sum-parton-2}),
we finally obtain
\begin{align}
\int dx \, b_1 (x) 
      = - \lim_{t \to 0} \frac{5}{24} t F_Q (t)
    + \frac{1}{9} \int dx
      \, \left [ \, 4 \, \delta_{_T} \bar u (x) +  \delta_{_T} \bar d (x) 
                     +   \delta_{_T} \bar s (x)  \, \right ]_D .
\label{eqn:b1-sum-parton-3}
\end{align}

We should note that the derived sum rule is not a rigorous one, 
but it is based on the parton model.
The situation is the same as the Gottfried sum rule.
There is a similarity between these sum rules:
\begin{alignat}{2}
\int dx \, b_1 (x) 
   & = 0 &
   & + \frac{1}{9} \int dx
      \, \left [ \, 4 \, \delta_{_T} \bar u (x) +  \delta_{_T} \bar d (x) 
                     +   \delta_{_T} \bar s (x)  \, \right ] ,
\nonumber \\
\int \frac{dx}{x} \, [F_2^p (x) - F_2^n (x) ]
     & =  \frac{1}{3} &
 & +\frac{2}{3} \int dx \, [ \bar u(x) - \bar d(x) ] .
\label{eqn:b1-sum-gottfried}
\end{alignat}
The factor of 1/3 in the Gottfried sum rule comes from
the flavor dependence of the valence quark distributions:
$\int [ u_v(x) - d_v (x) ]/3=1/3$. In the same way,
the factor $ -  \lim_{t \to 0} (5/24) t F_Q (t) =0$
comes from the tensor-polarized valence quark distributions.
The difference of zero and a finite number in both sum rules
could be summarized as
{\it the valence-quark number depends on flavor, whereas
the number does not depend on the tensor polarization}.
As the violation of the Gottfried sum rule led to
fruitful studies of flavor asymmetric antiquark
distributions $\bar u - \bar d$ and its physics origins \cite{flavor3},
there is a good possibility that a finite value of the $b_1$ sum
indicates the tensor-polarized antiquark distributions
according to Eq. (\ref{eqn:b1-sum-gottfried}).
There were HERMES measurements on $b_1$ and it indicated
\cite{hermes05}
\begin{align}
\int_{0.002}^{0.85} dx b_1(x) & =
   [1.05 \pm 0.34 \text{ (stat)} \pm 0.35 \text{ (sys)}] \times 10^{-2} ,
\nonumber \\
\int_{0.02}^{0.85} dx b_1(x) & =
   [0.35 \pm 0.10 \text{ (stat)} \pm 0.18 \text{ (sys)}] \times 10^{-2} 
   \ \ \text{for the range} \ Q^2>1 \, \text{GeV}^2 ,
\label{eqn:hermes-b1-sum}
\end{align}
which suggests a finite tensor polarization for antiquarks.
It is very interesting to find the physics origin 
of finite tensor polarized distributions.
It will be tested by the approved JLab experiment on $b_1$ \cite{Jlab-b1}.

Both $b_1$ and Gottfried sum rules are obtained in the parton model,
and they provide useful guidelines for the tensor-polarized antiquark 
distributions $(\delta_{_T} \bar q)$ and the light-quark 
flavor dependence of the unpolarized antiquark distributions
$(\bar u/\bar d \,)$. However, one should be careful that they are not
rigorous sum rules. For example, the $\bar u (x)$ and $\bar d (x)$
distributions are assumed to be equal at very small $x$ according to 
current parametrizations on the unpolarized PDFs. Therefore,
the integral $\int dx (\bar u -\bar d)$ seems to converge at this stage,
but it could become infinite if a slight difference exists
between $\bar u(x)$ and $\bar d(x)$ at small $x$.
In the same way, the integral $\int \delta_{_T} \bar q (x)$ may not
be finite depending on the $x$ dependence of $\delta_{_T} \bar q (x)$
or on its $Q^2$ dependence.
As noted in recent studies of Ref. \cite{miller-b1}, the sum rule are 
not satisfied in some theoretical models. 
In a convolution model of deuteron structure function,
the $b_1$ sum is given by the product of a moment of lightcone momentum
distribution of the nucleon with a moment of unpolarized quark
distribution. The first nucleon part identically vanishes 
but the second PDF part could diverge at large $Q^2$, 
although it converges in some PDF parametrization 
at small $Q^2$, so that the whole integral is not certain.
We need more theoretical efforts on an appropriate description
of the tensor-polarized distributions, especially 
on the antiquark part, together with experimental measurements.

\section{Parametrization of tensor-polarized distributions}
\label{parametrization}

\vspace{0.2cm}

Theoretical models are used for calculating structure functions,
and there are studies on low moments of the structure functions
by lattice QCD. However, it is almost impossible to obtain
$x$-dependent distributions at this stage,
so that global analyses of world experimental
data are used for determining reliable PDFs.
In the same way, a useful parametrization of the tensor polarized
distributions can be proposed by using the existing HERMES data,
although the number of data is not sufficient for an accurate
determination. It should be useful for proposing future experiments
({\it e.g.} \cite{Jlab-b1}) and for testing theoretical calculations.

We consider that certain fractions of the unpolarized distributions
are tensor polarized:
\begin{align}
\delta_{_T} q_{iv}^D (x) = \delta_{_T} w(x) \, q_{iv}^D (x), 
\ \ \ 
\delta_{_T} \bar q_i^D (x) 
           = \alpha_{\bar q} \, \delta_{_T} w(x) \, \bar q_i^D (x) ,
\label{eqn:delta-q-qbar-1}
\end{align}
where $\delta_{_T} w(x)$ and $\alpha_{\bar q} \, \delta_{_T} w(x)$
are such fractions for valence quarks and antiquarks, respectively.
The $x$ dependence of $\delta_{_T} w(x)$ could be different between
valence quarks and antiquarks, and flavor dependence may exist.
However, it is not the stage to investigate such details because
the HERMES data are the only ones. 
The tensor-polarized distributions cannot be determined
accurately at this stage, so that simplifying assumptions are
employed for the unpolarized PDFs.
Nuclear modifications in the PDFs of the deuteron are considered
as a few percent effects \cite{hkn07}, and they are neglected in our studies.
Namely, the PDFs of the deuteron are given by the contributions
from proton and neutron: $q_i^D=(q_i^{\,p}+q_i^{\,n})/2$ and 
$\bar q_i^D=(\bar q_i^{\,p}+\bar q_i^{\,n})/2$.
Next, isospin symmetry is used for relating the PDFs
of the neutron to the ones of the proton:
$u_n=d$, $d_n=u$, $\bar u_n=\bar d$, and $\bar d_n=\bar u$,
and flavor symmetric tensor-polarized distributions are assumed.
Then, we have the distributions
\begin{align}
\delta_{_T} q_v^D (x) & \equiv \delta_{_T} u_v^D (x) 
                         = \delta_{_T} d_v^D (x) 
          = \delta_{_T} w(x) \, \frac{u_v (x) +d_v (x)}{2} , 
\nonumber \\
\delta_{_T} \bar q^D (x) & \equiv \delta_{_T} \bar u^D (x)
                            = \delta_{_T} \bar d^D (x)
                            = \delta_{_T}      s^D (x)
                            = \delta_{_T} \bar s^D (x)                    
\nonumber \\
\    & = \alpha_{\bar q} \, \delta_{_T} w(x) \, 
    \frac{2 \bar u(x) +2 \bar d(x) +s(x) + \bar s(x)}{6} ,
\label{eqn:dw(x)}
\end{align}
for an analysis of the HERMES data.
Any available unpolarized PDFs could be used, but
the LO version of the MSTW parametrization \cite{mstw08} is employed.
From these tensor-polarized distributions, we finally obtain
$b_1$ for the deuteron as
\begin{align}
b_1^D (x) & = \frac{1}{36} \delta_{_T} w(x) \, \left [ \,
     5 \{ u_v (x) + d_v (x) \}   
  +4 \alpha_{\bar q} \{ 2 \bar u (x) + 2 \bar d (x)
  +   s (x) + \bar s (x) \} \, \right ] .
\label{eqn:b1x}
\end{align}
The important point of the analysis is how to choose
the $x$ dependence of $\delta_{_T} w(x)$. From the derivation
of the $b_1$ sum rule, the tensor-polarized valence-quark
distributions should satisfy the sum $\int dx (b_1)_{\text{valence}}=0$.
In order to satisfy this relation, the function $\delta_{_T} w(x)$
should have a node at least, so we may take the parametrization
\begin{equation}
\delta_{_T} w(x) = a x^b (1-x)^c (x_0-x) ,
\label{eqn:dw(x)-abc}
\end{equation}
where $a$, $b$, $c$, and $x_0$ are the parameters determined by the analysis.
The existence of the node is also supported by the convolution model
for $b_1$ with a D-state admixture. However, if the first moment 
of $\delta_{_T} w(x)$ vanishes, $x_0$ can be expressed by other ones as
\begin{equation}
x_0= \frac{\int_0^1 dx x^{b+1} (1-x)^c 
          \{ u_v (x) + d_v (x) \} }
          {\int_0^1 dx x^{b} (1-x)^c 
          \{ u_v (x) + d_v (x) \} } .
\end{equation}
There is no data to probe the scaling violation at this stage,
so that the $Q^2$ dependence is neglected in this analysis.
For calculating the unpolarized PDFs Eq. (\ref{eqn:b1x}), 
the average $Q^2$ value ($Q^2$=2.5 GeV$^2$) 
of the HERMES experiment is used.

\begin{table*}[t]
\caption{Determined parameters 
are listed for two sets of analyses.}
\label{parameter-delta-q}
\centering
\begin{tabular}{@{\hspace{0.2cm}}c|@{\hspace{0.4cm}}c@{\hspace{0.4cm}}
c@{\hspace{0.4cm}}c@{\hspace{0.4cm}}c@{\hspace{0.4cm}}
c@{\hspace{0.4cm}}c@{\hspace{0.2cm}}}
\hline
Analysis  &  $\chi^2$/d.o.f.    &  $a$                
          &  $\alpha_{\bar q}$  &  $b$  
          &  $c$                &  $x_0$              \\
\hline
Set 1     &  2.83               &  0.378 $\pm$ 0.212     
          &  0.0 (fixed)        &  0.706 $\pm$ 0.324 
          &  1.0  (fixed)       &  0.229              \\
Set 2     &  1.57               &  0.221 $\pm$ 0.174 
          &  3.20 $\pm$ 2.75    &  0.648 $\pm$ 0.342
          &  1.0  (fixed)       &  0.221              \\
\hline
\end{tabular}
\end{table*}

In order to find the impact of the tensor-polarized antiquark distributions,
we made two types of analyses:
\begin{itemize}
\item Set 1: Tensor-polarized antiquark distributions are terminated
             ($\alpha_{\bar q} = 0$).
\item Set 2: Finite tensor-polarized antiquark distributions
             are allowed ($\alpha_{\bar q}$ is a parameter).
\end{itemize}
The constant $c$ is fixed at $c=1$ in these analyses due to 
the lack of data at large $x$ to constrain it.
The obtained parameters are listed in Table \ref{parameter-delta-q}.
The smaller $\chi^2$ value of the set-2 analysis indicates
that there is a significant improvement in the fit by
including the tensor-polarized antiquark distributions.

\begin{figure}[h!]
\parbox[t]{0.48\textwidth}{
   \begin{center}
   \vspace{0.0cm}
    \epsfig{file=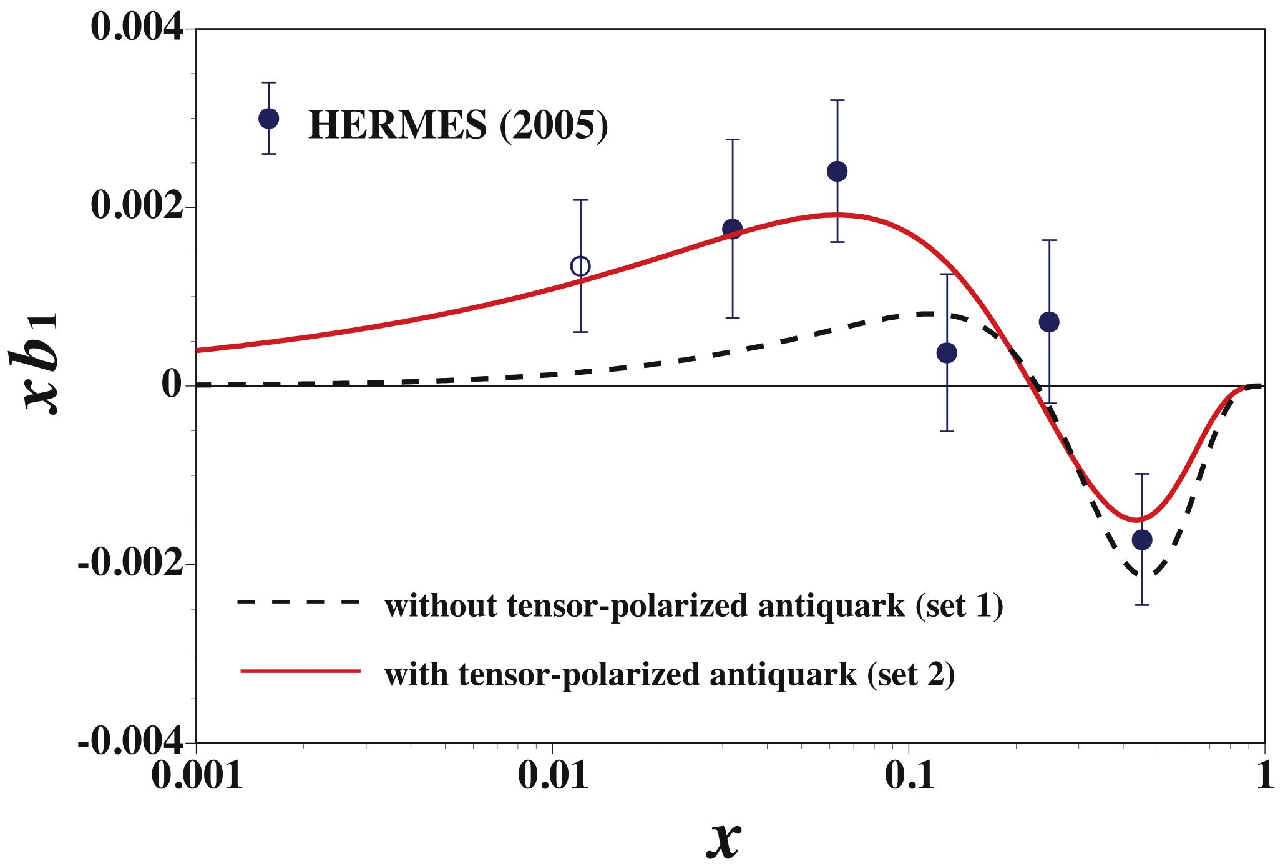,width=0.45\textwidth}
   \end{center}
\vspace{-0.4cm}
\caption{Obtained $b_1$ structure functions are compared with HERMES data.}
\label{fig:hermes-data}
}\hfill
\parbox[t]{0.48\textwidth}{
   \begin{center}
    \epsfig{file=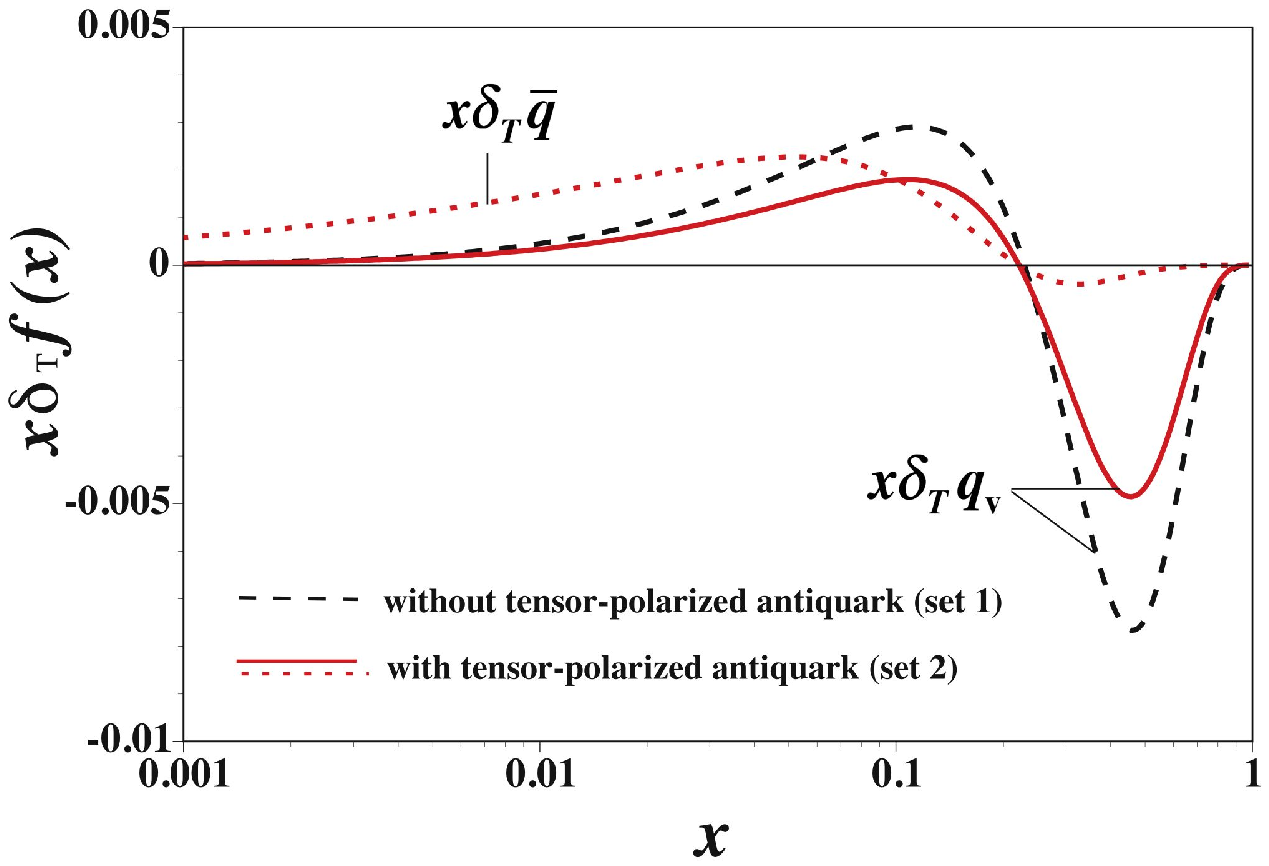,width=0.45\textwidth}
   \end{center}
\vspace{-0.4cm}
\caption{Determined tensor polarized distributions are shown.}
\label{fig:tensor-distribution}
}
\end{figure}

Obtained $b_1$ structure functions are compared with the HERMES data
in Fig.\,\ref{fig:hermes-data}. The dashed and solid curves are 
for the set-1 and the set-2,
respectively. A reasonable fit was obtained by the set-2 analysis,
whereas the small-$x$ ($<0.1$) data cannot be explained by the set-1.
Namely, the HERMES data at small $x$ suggest the existence of
a finite tensor-polarized antiquark distributions according 
to this analysis. The medium and large $x$ regions are well
explained by the tensor-polarized valence-quark distributions.

Determined tensor polarized distributions are shown in 
Fig.\,\ref{fig:tensor-distribution}.
The dashed and solid curves are tensor-polarized valence-quark 
distributions for the set-1 and set-2, respectively, and 
the dotted curve indicates the tensor-polarized antiquark
distribution of the set-2. 
The valence-quark distribution is negative at medium $x$ and 
it turned to negative at $x<0.2$. The antiquark distribution
becomes large at $x<0.2$. The functional form of
the valence-quark distribution is expected in
the convolution description of the deuteron structure function
\cite{kh91}. However, the physics origin of the antiquark distribution
is not obvious at this stage. More theoretical efforts are needed
to understand the origin.
From the obtained distributions of the set-2, 
the $b_1$ sum is estimated as
\begin{align}
\int dx \, b_1 (x) & = - \frac{5}{24} \lim_{t \rightarrow 0} t F_Q (t)
  + \frac{1}{18} \int dx \, [ \, 8 \delta_{_T} \bar u (x) 
    + 2  \delta_{_T} \bar d(x) +  \delta_{_T} s (x) 
    +  \delta_{_T} \bar s(x) \, ]  
\nonumber \\
& = 0.0058 .
\label{eqn:b1x-sum-value}               
\end{align}
This finite value is due to the existence of antiquark
tensor polarization.
There will be measurements on $b_1$ at JLab \cite{Jlab-b1},
so that much details will become clear experimentally. 
On the other hand, theoretical studies are needed
for the tensor-polarized quark and antiquark distributions.

\vspace{-0.0cm}
\noindent
\vspace{-0.55cm}

\section{Polarized proton-deuteron Drell-Yan processes}
\label{pd-drell-yan}

\vspace{0.2cm}
\subsection{Structure functions and spin asymmetries}

\vspace{0.2cm}

The HERMES data indicated an existence of finite tensor-polarized 
antiquark distributions. One of methods to probe
antiquark distributions is to use Drell-Yan processes. For example,
the polarized proton-deuteron (pd) Drell-Yan processes could be used with
tensor-polarized deuteron. There is no experiment for the pd Drell-Yan,
and there are only a few theoretical formalisms \cite{pd-drell-yan}.

We consider the Drell-Yan process of
$A \, (\text{spin} \, 1/2) + B \, (\text{spin} \, 1) \rightarrow \ell^+ \ell^- +X$.
Our formalism can be used for any spin-1/2 and spin-1 hadrons, but
the most realistic reaction is the proton-deuteron Drell-Yan.
Its cross section is written in terms of the lepton tensor
$L_{\mu\nu}$ and the hadron tensor $W^{\mu\nu}$
\begin{equation}
\frac{d\sigma}{d^4 Q \, d \Omega} = \frac{\alpha^2}{2 \, s \, Q^4} 
                                    \, L_{\mu\nu} \, W^{\mu\nu}
\ ,
\label{eqn:cross-0}
\end{equation}
where $\alpha=e^2/(4\pi)$ is the fine structure constant,
$s$ is the center-of-mass energy squared $s=(P_A+P_B)^2$,
$Q$ is the total dilepton momentum, and $\Omega$ is the solid
angle of the momentum $\vec k_{\ell^+}-\vec k_{\ell^-}$.

A general spin-density formalism, by expressing structure functions
in terms of helicity amplitudes and Clebash-Gordan coefficients
with the conditions of Hermiticity, parity conservation, 
and time-reversal invariance, indicates that
there exist 108 structure functions for the unpolarized and polarized
pd Drell-Yan processes, whereas there are 48 structure functions for the pp.
There are 60 new structure functions, which should be associated with
the deuteron tensor structure. Of course, all of them are not important
for the first investigation. 
If the cross section is integrated over the lepton-pair transverse
momentum $\vec Q_T$, there are only 22 ones which include 11 new functions 
associated with the tensor structure of the deuteron. 
This spin-density formalism is rather lengthy, so that we refer
the paper \cite{pd-drell-yan} for the details.

\begin{figure}[t]
   \begin{center}
       \epsfig{file=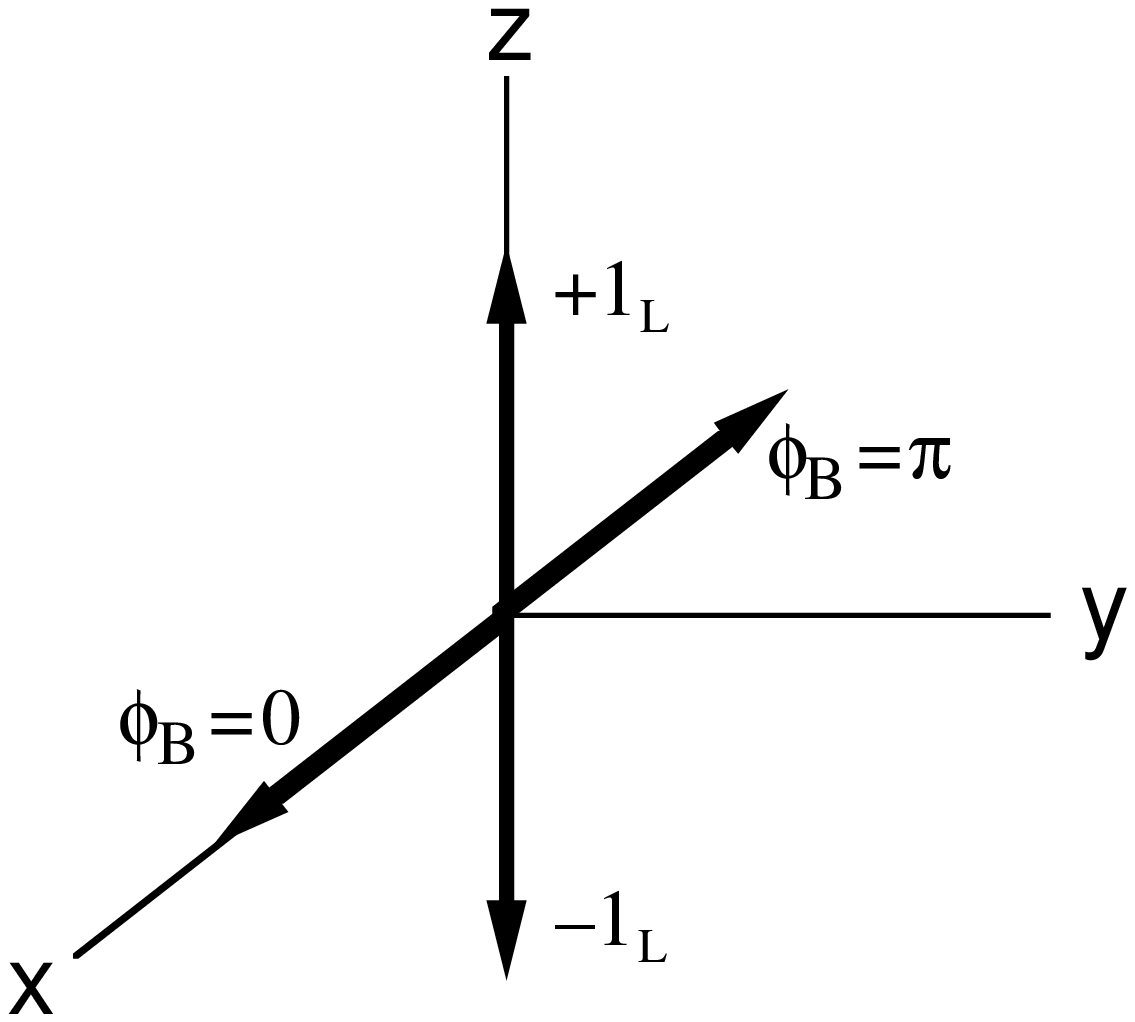,width=4.3cm}
       \epsfig{file=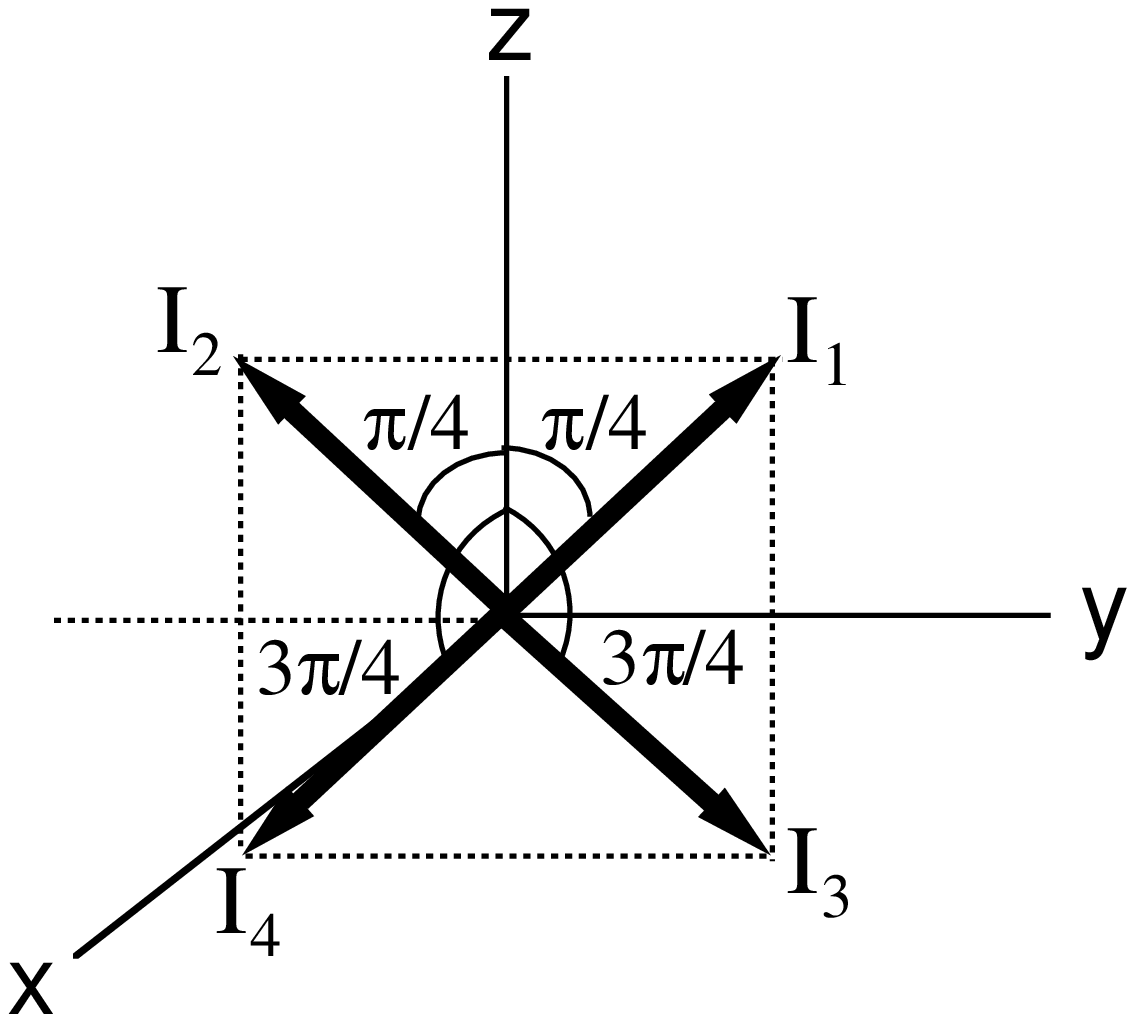,width=4.3cm}
       \epsfig{file=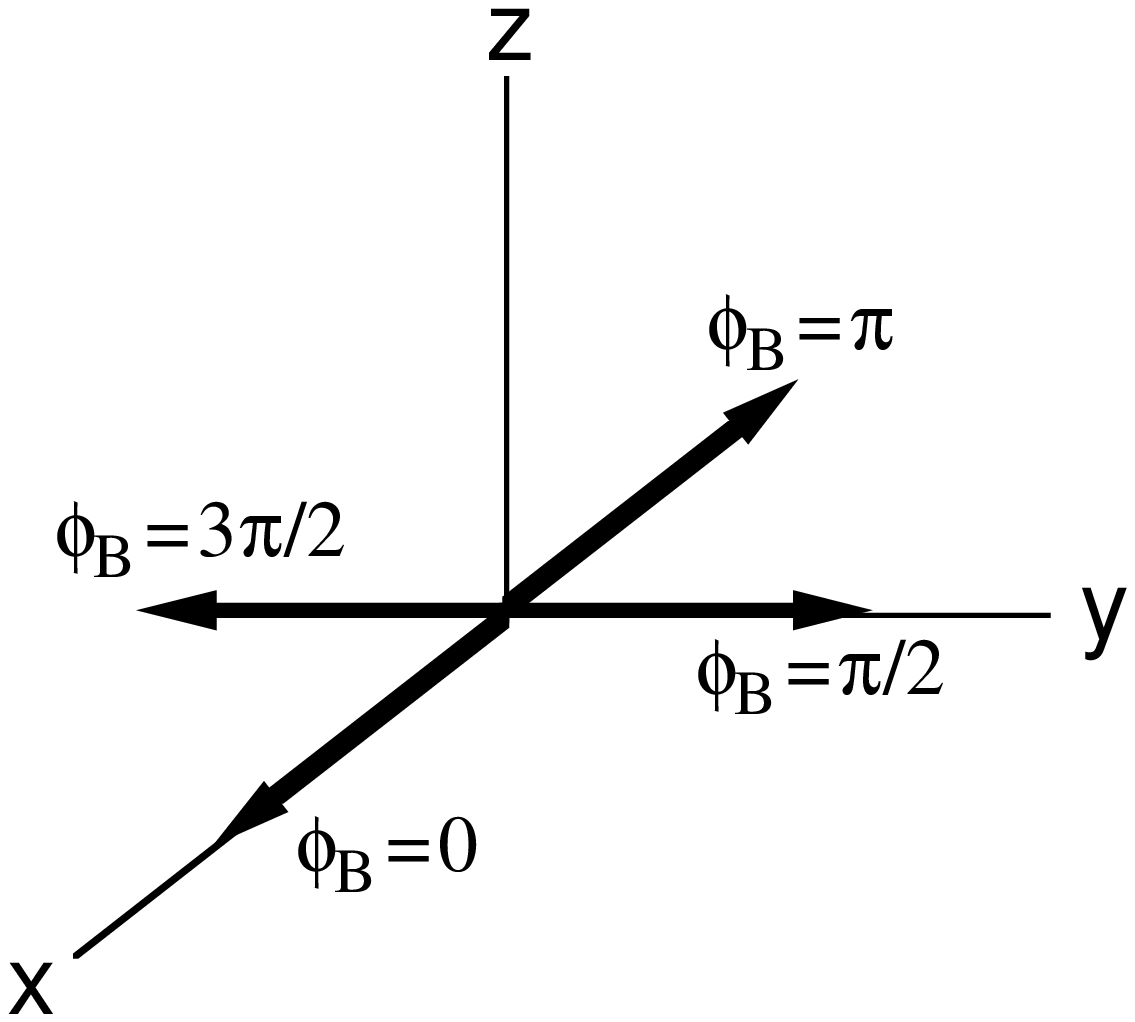,width=4.3cm}
   \end{center}
\vspace{-0.5cm}\hspace{2.7cm}{$(a) \, Q_0$}
\hspace{3.3cm}{$(b) \, Q_1$}
\hspace{3.3cm}{$(c) \, Q_2$}
\vspace{-0.0cm}
\caption{Tensor polarizations $Q_0$, $Q_1$, and $Q_2$.}
\label{fig:q012}
\end{figure}

In the general hadron tensor formalism, 
the hadron tensor of the Drell-Yan process
\begin{equation}
W^{\mu \nu} =  \int 
     \frac{d^4 \xi}{(2 \, \pi)^4} \, e^{i Q\cdot \xi} \,
 \langle \, P_A S_A P_B S_B \, | \, J^\mu (0) J^\nu (\xi) 
        \, | \, P_A S_A P_B S_B \, \rangle
 ,
\end{equation}
is expanded in terms of possible Lorentz index combinations including
the hadron momenta and spins by considering the conditions
\begin{alignat}{2}
& \text{Hermiticity:} \ \ &
[W^{\nu \mu}(Q; P_A S_A; P_B S_B)]^* 
           &  = W^{\mu \nu} (Q; P_A S_A; P_B S_B)  ,
\nonumber \\  
& \text{Parity conservation:} \ \ &
W^{\mu \nu}(Q; P_A S_A; P_B S_B) & =
  W_{\mu \nu}(\overline{Q}; \overline{P}_A -\overline{S}_A; 
                            \overline{P}_B -\overline{S}_B)  ,
\nonumber \\  
&  \text{Time-reveral invariance:} \ \ &
[W^{\mu \nu}(Q; P_A S_A; P_B S_B)]^*  & = 
         W_{\mu \nu}(\overline{Q}; \overline{P}_A \overline{S}_A; 
                                   \overline{P}_B \overline{S}_B) ,
\label{eqn:hpt}
\end{alignat}
where $\overline P$ is defined by $\overline P^{\, \mu} = (P^0, -\vec P)$.
For the expansion, the Lorentz vectors $X^\mu$, $Y^\mu$, and $Z^\mu$
\begin{align}
X^{\mu}  & =  P_A^{\mu} \, Q^2 \, Z \cdot P_B 
            - P_B^{\mu} \, Q^2 \, Z \cdot P_A
   + Q^{\mu} \, (Q \cdot P_B \, Z \cdot P_A
                          - Q\cdot P_A \, Z \cdot P_B ) ,
\nonumber \\
Y^{\mu} & = \epsilon^{\mu \alpha \beta \gamma} 
               \ P_{A \alpha} \, P_{B \beta} \, Q_{\gamma}
, \ \ \ 
Z^{\mu}  = P_{A}^{\mu} \, Q \cdot P_{B}- P_{B}^{\mu} \, Q \cdot P_A
,
\label{eqn:xyz}
\end{align}
are used. We also define the vector $T^\mu$ by
\begin{equation}
T^\mu = \varepsilon^{\mu\alpha\beta\gamma} S_{\alpha} \, Z_\beta \, Q_\gamma
\, .
\end{equation}
In addition to these vectors, $Q^\mu$, 
$S_A^\mu$, $S_B^\mu$, $T_A^\mu$, and $T_B^\mu$ 
can be used to expand $W^{\mu\nu}$.
However, instead of $S_A^\mu$ and $S_B^\mu$, it is more convenient 
to use the transverse vectors $S_{AT}^{\mu}$ and $S_{BT}^{\mu}$:
\begin{equation}
S_T^{\mu} = \left ( g^{\mu\nu} - \frac{Q^\mu Q^\nu}{Q^2}
                               - \frac{Z^\mu Z^\nu}{Z^2} \right ) S_\nu \, .
\end{equation}
In the general case, it is too lengthy to write them down here
becasuse there are 108 structure functions. The transverse momentum
$Q_T$ is roughly restricted by the hadron size $R$ by $Q_T < 1/R$,
so that the limit of $Q_T\rightarrow 0$ is considered in the following formalism.
Then, $X^\mu$ and $Y^\mu$ do not have to be considered because
$|\vec X|$ and $|\vec Y|$ are proportional to $Q_T$ and $X^0=Y^0=0$
in the dilepton rest frame.
Then, we obtain
\begin{align}
W ^{\mu\nu} = & - g^{\mu\nu} A - \frac{Z^\mu Z^\nu}{Z^2} B
        + Z^{ \{ \mu} T_A^{\nu \} } C + Z^{ \{ \mu} T_B^{\nu \} } D
   + Z^{ \{ \mu} S_{AT}^{\nu \} } E  + Z^{ \{ \mu} S_{BT}^{\nu \} }  F 
\nonumber \\
&  - S_{BT}^{\mu} S_{BT}^{\nu}    G - S_{AT}^{ \{ \mu} S_{BT}^{\nu \} } H
   + T_A^{ \{ \mu} S_{BT}^{\nu \} } I + S_{BT}^{\{ \mu} T_B^{\nu \} } J 
   + Q^\mu Q^\nu K + Q^{ \{ \mu} Z^{\nu \} } L 
\nonumber \\
&  + Q^{ \{ \mu} S_{AT}^{\nu \} } M   + Q^{ \{ \mu} S_{BT}^{\nu \} } N 
   + Q^{ \{ \mu} T_A^{\nu \} } O   + Q^{ \{ \mu} T_B^{\nu \} } P
\ ,
\label{eqn:w1}
\end{align}
where $Q^{ \{ \mu} Z^{\nu \} }$ is defined by
$ Q^{ \{ \mu} Z^{\nu \} } \equiv Q^\mu Z^\nu + Q^\nu Z^\mu $.
Next, we impose the current conservation $Q_\mu W^{\mu\nu}=0$,
and then the coefficients $A$, $B$, $\cdots$, which still
contain spin factors,
are expanded by the scalar and pseudoscalar terms with the spins.
We finally obtain
\begin{align}
W^{\mu \nu} =& -\left[ g^{\mu \nu} - \frac{Q^{\mu} Q^{\nu}}{Q^2} \right] 
           \left\{W_{0,0} + \frac{M_A M_B}{s \, Z^2} \,
                        Z \cdot S_A \, Z \cdot S_B \, V_{0,0}^{LL} 
                       - S_{AT} \cdot S_{BT} \, V_{0,0}^{TT} \right.  
\nonumber \\ 
   & \ \ \ \ \ \ \ \ \ \ \ \ 
  - \left(\frac{8 \, M_B^2 \, ( Z \cdot S_B )^2}{s^2 \, ( Q \cdot P_B )^2} 
                             + \frac{4}{3} \, S_B^2 \right) V_{0,0}^{U Q_0} 
                      + \frac{M_B}{Z^2 \, Q \cdot P_B} \,
                              Z \cdot S_B \, T_{A}\cdot S_{BT} \,
                               V_{0,0}^{TQ_1} \biggr\}  
\nonumber \\
                    & -\left[\frac{Z^{\mu} Z^{\nu}}{Z^2}
                      - \frac{1}{3} \left(g^{\mu \nu} 
                         - \frac{Q^{\mu} Q^{\nu}}{Q^2}\right) \right] 
           \left\{W_{2,0} + \frac{M_A M_B}{s \, Z^2} \,
                    Z \cdot S_A \, Z \cdot S_B \, V_{2,0}^{LL}  
                    - S_{AT} \cdot S_{BT} \, V_{2,0}^{TT} \right.  
\nonumber \\
   & \ \ \ \ \ \ \ \ \ \ \ \ 
                     - \left(\frac{8 \, M_B^2 \, ( Z \cdot S_B )^2}
                                {s^2 \, ( Q \cdot P_B )^2} 
                               + \frac{4}{3} \, S_B^2 \right) V_{2,0}^{U Q_0} 
                     + \frac{M_B}{Z^2 \, Q \cdot P_B} \, 
                             Z \cdot S_B \, T_{A} \cdot S_{BT} \, 
                                V_{2,0}^{TQ_1} \biggr\} 
\nonumber \\
                &- Z^{\{\mu} T_{A}^{\nu\}} \frac{1}{\sqrt{Q^2} \, Z^2} 
                     \left\{ U_{2,1}^{T U}
                  - \left( \frac{8 \, M_B^2 \, ( Z \cdot S_B )^2}
                      {s^2 \, ( Q \cdot P_B )^2} 
                        + \frac{4}{3}\, S_B^2 \right) U_{2,1}^{TQ_0} \right\} 
\nonumber \\
                &- Z^{\{\mu} T_{B}^{\nu\}} \frac{1}{\sqrt{Q^2} \, Z^2} 
                      \left\{ U_{2,1}^{U T}
                      + \frac{M_A M_B}{s \, Z^2} \, 
                 Z \cdot S_A  \, Z \cdot S_B \, U_{2,1}^{LQ_1} 
                 + S_{AT} \cdot S_{BT} \, U_{2,1}^{TQ_2} \right\} 
\nonumber \\
                   &+ Z^{\{\mu} S_{AT}^{\nu\}} \,
                      \frac{\sqrt{Q^2} \, M_B}{Z^2 \, Q \cdot P_B} \,
                      Z \cdot S_B \, U_{2,1}^{TL} 
\nonumber \\
                   &+ Z^{\{\mu} S_{BT}^{\nu\}} 
                    \left\{ - \frac{\sqrt{Q^2} \, M_A}{Z^2 \, Q \cdot P_A} \,
                      Z \cdot S_A \, U_{2,1}^{LT}
                      + \frac{\sqrt{Q^2} \, M_B}{Z^2 \, Q \cdot P_B} \, 
                        Z \cdot S_B \, U_{2,1}^{U Q_1} 
                   - \frac{1}{\sqrt{Q^2} \, Z^2} \, T_{A} \cdot S_{BT} \,
                      U_{2, 1}^{TQ_2}\right\} 
\nonumber \\
                &- \left[ 2 \, S_{BT}^{\mu} S_{BT}^{\nu} 
                      -  S_{BT}^2 \left( g^{\mu \nu} 
                               - \frac{Q^{\mu} Q^{\nu}}{Q^2} 
                               -\frac{Z^{\mu} Z^{\nu}}{Z^2} \right) \right]
                     U_{2,2}^{U Q_2}  
\nonumber \\
                &- \left[S_{AT}^{\{\mu} S_{BT}^{\nu\}} 
                    - S_{AT} \cdot S_{BT} \left( g^{\mu \nu} 
                              - \frac{Q^{\mu} Q^{\nu}}{Q^2} 
                              -\frac{Z^{\mu} Z^{\nu}}{Z^2} \right) \right] 
                    U_{2,2}^{TT} 
\nonumber \\
                &- \left[T_{A}^{\{\mu} S_{BT}^{\nu\}} 
                      - T_{A} \cdot S_{BT} \left( g^{\mu \nu} 
                               - \frac{Q^{\mu} Q^{\nu}}{Q^2} 
                               -\frac{Z^{\mu} Z^{\nu}}{Z^2} \right) \right] 
               \frac{M_B}{Z^2 \, Q \cdot P_B} \, Z \cdot S_B \,
                               U_{2,2}^{TQ_1} 
\nonumber \\
                &+ S_{BT}^{\{\mu} T_{B}^{\nu\}} \,
                  \frac{M_A}{Z^2 \, Q \cdot P_A} \, Z \cdot S_A \,
                               U_{2,2}^{LQ_2}
\ .
\end{align}
Therefore, in the limit of 
$Q_T\rightarrow 0$, threre are 22 structure functions:
\begin{alignat}{11}
& W_{0,0} \, ,  
               & & V_{0,0}^{LL} ,    & & V_{0,0}^{TT} , 
               & & V_{0,0}^{U Q_0} , & & V_{0,0}^{TQ_1} , 
               & & W_{2,0} ,         & & V_{2,0}^{LL} ,
               & & V_{2,0}^{TT} ,    & & V_{2,0}^{U Q_0} ,
               & & V_{2,0}^{TQ_1} ,  & & U_{2,1}^{T U} ,  
\nonumber \\
& U_{2,1}^{TQ_0} , \ \ \     
               & & U_{2,1}^{U T} ,  \ \ 
               & & U_{2,1}^{LQ_1} , \ \ 
               & & U_{2,1}^{TQ_2} , \ \ 
               & & U_{2,1}^{TL} ,   \ \ 
               & & U_{2,1}^{LT}  ,  \ \ 
             & & U_{2,1}^{U Q_1}  , \ \ 
             & & U_{2,2}^{U Q_2} ,  \ \ 
             & & U_{2,2}^{TT}  ,    \ \ 
             & & U_{2,2}^{TQ_1} ,   \ \ 
             & & U_{2,2}^{LQ_2} ,
\label{eqn:sf}
\end{alignat}
where $W$, $V$, and $U$ are an unpolarized structure function,
a polarized one without the spin factors in the hadron tensor,
and a polarized one with the spin factor. 
The function $W_{L,M}$ is obtained by the integral
$\int d\Omega \,  Y_{LM} \, d\sigma/(d^4Q \,d\Omega) \propto W_{L,M}$
of the unpolarized reaction. 
The superscripts $U$, $L$, and $T$ show unpolarized,
longitudinally polarized, and transversely polarized states.
The quadrupole polarizations $Q_0$, $Q_1$, and $Q_2$ 
are associated with the spherical harmonics 
$Y_{20}$, $Y_{21}$, and $Y_{22}$ as shown
in Fig.\,\ref{fig:q012}.
They are the polarizations in the $xz$, $yz$, and $xy$ planes.
The structure functions with $Q_0$, $Q_1$, and $Q_2$ are specific
for the spin-1 deuteron.

In the pp Drell-Yan processes, the unpolarized,
longitudinal, and transverse combinations exist: $< \! \sigma \! >$,
$A_{LL}$, $A_{TT}$, $A_{LT}$, and $A_{T}$.
In addition, the following fifteen quadrupole spin asymmetries could
be investigated in the pd Drell-Yan:
\begin{alignat}{8}
& < \! \sigma \! >, \ \ & & 
A_{LL}, \ \             & &
A_{TT}, \ \             & &
A_{LT}, \ \             & &
A_{TL}, \ \             & &
A_{UT}, \ \             & &
A_{TU}, \ \             & &
        \ \        \nonumber \\
& A_{UQ_0}, \ \         & &     
A_{TQ_0}, \ \           & &
A_{UQ_1}, \ \           & &
A_{LQ_1}, \ \           & &
A_{TQ_1}, \ \           & &
A_{UQ_2}, \ \           & &
A_{LQ_2}, \ \           & &
A_{TQ_2}.
\end{alignat}
These asymmetries are expressed in terms of the structure functions
in Eq. (\ref{eqn:sf}). For example, the quadrupole spin asymmetry $A_{UQ_0}$
is measured with the unpolarized proton and the $Q_0$-type
tensor polarized deuteron, and it is expressed in terms of
the structure functions $V_{0,0}^{UQ_0}$,
$V_{2,0}^{UQ_0}$, $W_{0,0}$, and $W_{2,0}$:
\begin{equation}
\! \! \! \! \! \! \! \! \! \! \! \! \! \! \! 
A_{UQ_0} = \frac{1}{2 < \! \sigma \! >} \,        
         \bigg [ \, \sigma(\bullet , 0_L)
            - \frac{ \sigma(\bullet , +1_L) 
                    +\sigma(\bullet , -1_L) }{2} \, \bigg ]   
         =  \frac{ 2 \, V_{0,0}^{UQ_0} 
                          + (\frac{1}{3}-cos^2 \theta ) \, 
                            V_{2,0}^{UQ_0} }
                 { 2 \, W_{0,0}
                     + (\frac{1}{3}-cos^2 \theta ) \,  W_{2,0}  }
\, ,
\label{eqn:a-uq0}
\end{equation}
where $\bullet$ indicates the unpolarized case.

\subsection{Parton model expressions}

\vspace{0.2cm}

\begin{wrapfigure}[10]{r}{0.42\textwidth}
   \vspace{-0.7cm}
   \begin{center}
       \epsfig{file=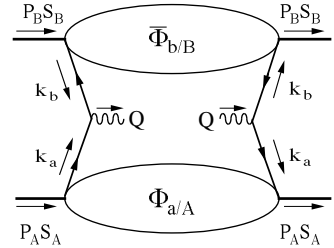,width=4.2cm}
   \end{center}
   \vspace{-0.5cm}
   \caption{Paton model for Drell-Yan.}
       \label{fig:parton}
\end{wrapfigure}

Possible structure functions and spin asymmetries were introduced
for the pd Drell-Yan processes. Here, we express them in terms of
parton distribution functions of the proton and deuteron \cite{pd-drell-yan}.
As shown in Fig.\,\ref{fig:parton}.
The hadron tensor $W^{\mu\nu}$ of the Drell-Yan processes is written
by the quark and antiquark correlation functions, which are expressed
by combinations of possible vectors and pseudovectors.
The leading contribution to the hadron tensor is
\begin{equation}
W^{\mu \nu} = \frac{1}{3} \sum_{a, b} \delta_{b \bar{a}} \, e_a^2 
            \int d^4 k_a \, d^4 k_b \, \delta^4 (k_a + k_b - Q) \, 
            Tr [\Phi_{a/A} (P_A S_A; k_a) \gamma^\mu
           \bar{\Phi}_{b/B} (P_B S_B; k_b) \gamma^\nu]
\, .
\label{eqn:w-2}
\end{equation}
The correlation functions $\Phi_{a/A}$ and $\bar \Phi_{\bar a/B}$ are
defined by
\begin{align}
\Phi_{a/A} (P_A S_A; k_a)_{ij} & =
           \int \frac{d^4 \xi}{(2 \pi)^4} \, e^{i k_a \cdot \xi}
\, \langle \, P_A S_A \, | \, 
\bar{\psi}_j^{(a)}(0) \, \psi_i^{(a)}(\xi) \, | \, P_A S_A
\, \rangle \,
\, ,
\nonumber \\
\bar{\Phi}_{\bar a/B} (P_B S_B; k_{\bar a})_{ij} & =
           \int \frac{d^4 \xi}{(2 \pi)^4} \, e^{i k_{\bar a} \cdot \xi}
\, \langle \, P_B S_B \, | \, \psi_i^{(a)}(0) \, \bar{\psi}_j^{(a)}(\xi) \,
 | \, P_B S_B
\, \rangle 
\, ,
\end{align}
where link operators for the gauge invariance are not explicitly written.
Using a Fierz transformation, we write the hadron tensor in
a factorized form:
\begin{align}
 W^{\mu \nu} = & \frac{1}{3} \sum_{a, b} \delta_{b \bar{a}} \, e_a^2 
      \int d^2 \vec{k}_{aT} \, d^2 \vec{k}_{bT} \,
      \delta^2 (\vec{k}_{aT} + \vec{k}_{bT} - \vec{Q}_T) \, 
      \bigg[ \, \bigg\{ \, 
    - \Phi_{a/A} [\gamma^\alpha] \, \bar{\Phi}_{b/B}[\gamma_\alpha]
\nonumber \\
&   - \Phi_{a/A} [\gamma^\alpha \gamma_5] \,
      \bar{\Phi}_{b/B}[\gamma_\alpha \gamma_5] 
    + \frac{1}{2} \Phi_{a/A} [i \sigma_{\alpha \beta} \gamma_5] \,
      \bar{\Phi}_{b/B}[i \sigma^{\alpha \beta} \gamma_5] \bigg\} \, g^{\mu\nu} 
    + \Phi_{a/A} [\gamma^{\{\mu}] \, \bar{\Phi}_{b/B}[\gamma^{\nu\}}]
\nonumber \\
&   + \Phi_{a/A} [\gamma^{\{\mu}\gamma_5] \,
      \bar{\Phi}_{b/B}[\gamma^{\nu\}}\gamma_5]
    + \Phi_{a/A} [i \sigma^{\alpha\{\mu}\gamma_5] \, 
      \bar{\Phi}_{b/B}[i \sigma_{\ \ \, \alpha}^{\nu\}}\gamma_5] \, \bigg] \, 
    + O(1/Q)
\ .
\label{eqn:w-3}
\end{align}
\begin{wrapfigure}[11]{r}{0.39\textwidth}
   \vspace{-0.9cm}
   \begin{center}
       \epsfig{file=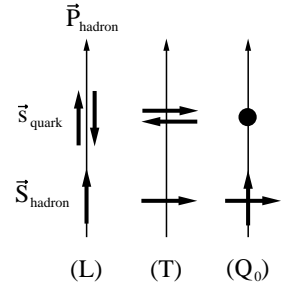,width=4.0cm}
   \end{center}
   \vspace{-0.7cm}
   \caption{Longitudinally polarized, transversity, 
              and tensor polarized distributions. 
              The notation $\bullet$ indicates unpolarized.}
       \label{fig:ltq}
\end{wrapfigure}

\noindent
Then, these correlation functions are expressed by 
the unpolarized, longitudinally-polarized, transversity
distributions of the proton and deuteron, together with
the tensor-polarized distributions of the deuteron,
as illustrated in Fig.\,\ref{fig:ltq}.
Particularly, it is important that the correlation function 
$\Phi [\gamma^\mu]$
contains the tensor-polarized distributions $\delta_{_T} q (x)$.
The details are found  in Ref. \cite{pd-drell-yan}.

In the naive parton model, we find 19 structure functions, which
become four by the $\vec Q_T$ integration.
We define $\overline W = \int d^2 \vec Q_T \, W$, and 
$\overline V$ and $\overline U$ are defined in the same way.
The pd Drell-Yan cross section is then given by
\begin{align}
\frac{d \sigma}{dx_A \, dx_B \, d \Omega} = \frac{\alpha^2}{4 \, Q^2} \,
      \bigg [ \,  & (1 + \cos^2 \theta) \bigg\{ \, 
            \overline W_T 
            + \frac{1}{4} \lambda_A \lambda_B \, \overline V_T^{\, LL}
            + \frac{2}{3}  \left( 2 \, |\vec S_{BT}|^2 - \lambda_B^2 \right)
                               \, \overline V_T^{\, UQ_0} \, \bigg\}
\nonumber \\
&  +  \sin^2 \theta \, |\vec S_{AT}| \, |\vec S_{BT}|
             \cos(2\phi-\phi_A-\phi_B) \, \overline U_{2,2}^{\, TT}
        \, \bigg ] \, .
\end{align}
The structure functions are expressed by the parton distributions
in the process $q$(in p)+$\bar q$(in d)$\rightarrow \ell^+ + \ell^-$ as
\begin{alignat}{2}
\overline W_T     & = \frac{1}{3} \sum_i e_i^2 \, 
                 q_i (x_A) \, \bar q_i (x_B) \, , \ \ \ \ \ \ 
& 
\overline V_T^{\, LL} & = - \frac{4}{3} \sum_i e_i^2 \, 
          \Delta q_i (x_A) \, \Delta \bar q_i (x_B) \, , 
\nonumber \\
\overline U_{2,2}^{\, TT} & = \frac{1}{3} \sum_i e_i^2 \, 
          \Delta_{_T} q_i (x_A) \, \Delta_{_T} \bar q_i (x_B)  \, , \ \ \ \ \ \ 
& 
\overline V_T^{\, UQ_0} & 
            = \frac{1}{6} \sum_i e_i^2 \, 
                          q_i (x_A) \, \delta_{_T} \bar q_i (x_B) 
\, ,
\end{alignat}
where $\Delta q_i$ and $\Delta_{_T} q_i$ are longitudinally-polarized
and transversity distributions.
The tensor-polarized distributions can be studied 
by the asymmetry $Q_0$:
\begin{equation}
A_{UQ_0} = \frac{\overline V_T^{\, UQ_0}}{\overline W_T}
         = \frac{\sum_a e_a^2 \, 
                  \left[ \, q_a(x_A) \, \delta_{_T} \bar q_a(x_B)
                     + \bar q_a(x_A) \, \delta_{_T} q_a(x_B) \, \right] }
                {2  \sum_a e_a^2 \, 
                  \left[ \, q_a(x_A) \, \bar q_a(x_B)
                          + \bar q_a(x_A) \, q_a(x_B) \, \right] }
\ .
\end{equation}
This asymmetry indicates that the tensor-polarized distributions
$\delta_{_T} q$ and $\delta_{_T} \bar q$ should be found 
in the Drell-Yan in addition to the charged-lepton scattering.
It is especially important that the antiqurk distribution 
$\delta_{_T} \bar q$ is measured, because it was suggested
that the finite $\delta_{_T} \bar q$ was indicated by
the HERMES experiment.

The possibility of polarized deuteron acceleration was once 
considered at RHIC \cite{rhic-d}, but it was not attained.
However,  there are future possibilities to investigate 
the Drell-Yan process with a fixed tensor-polarized deuteron target 
at hadron faclities such as Fermilab \cite{fermilab-d}, 
J-PARC \cite{j-parc}, GSI \cite{gsi-fair}, and CERN-COMPASS.
The JLab $b_1$ measurement will clarify the details of the tensor-polarized
distributions in 2020's. Together with their data, we expect that
the Drell-Yan measurements will clarify the tensor-polarized antiquark
distributions.

\section{Summary}
\label{summary}

\vspace{0.2cm}

We explained our studies on the tensor-polarized structure function $b_1$
and tensor-polarized quark and antiquark distributions,
$\delta_{_T} q$ and $\delta_{_T} \bar q$.
First, the projection operators are shown for all the eight
structure functions of the deuteron from the hadron tensor $W^{\mu\nu}$.
The projection operators should be useful in a convolution description
of the deuteron structure functions. 
Second, the sum rule was explained for $b_1$ by using the parton model.
It is valuable for indicating the existence of 
antiquark tensor polarization, as the Gottfried sum rule violation 
indicated a $\bar u/\bar d$ asymmetry in the nucleon.
Third, the parametrization of the tensor-polarized quark and 
antiquark distributions was proposed by analyzing the HERMES 
data on $b_1$. The analysis indicated an existence of 
antiquark tensor polarization, and its origin should be
investigated theoretically. The tensor-polarized antiquark distributions
should be studied by the Drell-Yan processes with tensor-polarized
deuteron. We showed the general formalisms for the structure 
functions in the proton-deuteron Drell-Yan processes and also
their expressions in terms of the parton distribution functions
in the proton and deuteron. With the tensor-polarized deuteron,
the Drell-Yan process probes the tensor-polarized antiquark
distributions directly. 
The JLab experiment on $b_1$ was approved and the actual measurement
is expected to start in 2019. There are other possibilities to investigate
the tensor structure at EIC, Fermilab, J-PARC, GSI, and CERN-COMPASS.
The studies of tensor-polarized structure
functions could open a new era of high-energy spin physics.




\section*{References}

\vspace{0.2cm}

\begin{thebibliography}{9}
\bibitem{fs83}  L. L. Frankfurt and M. I. Strikman, 
                   Nucl. Phys. A {\bf 405}, 557 (1983).
\bibitem{hjm89} P. Hoodbhoy, R. L. Jaffe, and A. Manohar,
                   Nucl. Phys. B {\bf 312}, 571 (1989);
        R. L. Jaffe and A. Manohar, Nucl. Phys. {\bf B321}, 343 (1989).
\bibitem{b1-sum} 
      F. E. Close and S. Kumano, Phys. Rev. D  {\bf 42}, 2377 (1990).
\bibitem{rho-production}
      A. Bacchetta and P. J. Mulders, Phys. Rev. D {\bf 62}, 114004 (2000). 
\bibitem{kh91} H. Khan and P. Hoodbhoy,
               Phys. Rev. C {\bf 44}, 1219 (1991).
\bibitem{miller-b1} 
      G. A. Miller, pp.30-33 in {\it Topical Conference on 
                Electronuclear physics with Internal Targets},
                edited by R. G. Arnold
                (World Scientific, Singapore, 1990).
      G. A. Miller, Phys. Rev. D {\bf 89}, 045203 (2014).
\bibitem{b1-shadowing}
      For example, see 
       N. N. Nikolaev and W. Sch\"afer,
                 Phys. Lett. B {\bf 398}, 245 (1997);
       Erratum, {\it ibid.}, B {\bf 407}, 453 (1997); 
       J. Edelmann, G. Piller, and W. Weise,
                 Z. Phys. A {\bf 357}, 129 (1997).
\bibitem{bj98}
      K. Bora and R. L. Jaffe, Phys. Rev. D {\bf 57}, 6906 (1998). 
\bibitem{spin-1-frag}
      A. Sch\"afer, L. Szymanowski, and O. V. Teryaev,
                  Phys. Lett. B {\bf 464}, 94 (1999). 
\bibitem{spin-1-gpd} 
      E. R. Berger, F. Cano, M. Diehl, and B. Pire, 
                  Phys. Rev. Lett. {\bf 87}, 142302 (2001);
      A. Kirchner and D. Mueller,
                  Eur. Phys. J. C {\bf 32}, 347 (2003); 
      M. Diehl, Phys. Rept. {\bf 388}, 41 (2003);
      F. Cano and B. Pire, Eur. Phys. J. A {\bf 19}, 423 (2004);
      A. V. Belitsky and A. V. Radyushkin, 
                Phys. Rept. {\bf 418}, 1 (2005). 
\bibitem{mass-corr} W. Detmold, Phys. Lett. B {\bf 632}, 261 (2006).
\bibitem{dmitrasinovic-96}
      V. Dmitrasinovic, Phys. Rev. D {\bf 54}, 1237 (1996).
\bibitem{lattice} C. Best {\it et al.}, 
                   Phys. Rev. D {\bf 56}, 2743 (1997).
\bibitem{kk08} T.-Y. Kimura and S. Kumano, 
                   Phys. Rev. D {\bf 78}, 117505 (2008).                              
\bibitem{angular-spin-1}	
     S. K. Taneja, K. Kathuria, S. Liuti, and G. R. Goldstein,
                Phys. Rev. D {\bf 86}, 036008 (2012).
\bibitem{pd-drell-yan}
      S. Hino and S. Kumano, Phys. Rev. D {\bf 59}, 094026 (1999);
                                          {\bf 60}, 054018 (1999);
      S. Kumano and M. Miyama, Phys. Lett. B {\bf 479}, 149 (2000). 
\bibitem{hermes05}
      A. Airapetian {\it et al.} (HERMES Collaboration), 
                   Phys. Rev. Lett. {\bf 95}, 242001 (2005);
      C. Riedl, talk at the
         Tensor Polarized Solid Target Workshop, March 10-12, 2014,
         JLab, Newport News, USA,
         http://www.jlab.org/conferences/tensor2014/.
\bibitem{Jlab-b1} Proposal to Jefferson Lab PAC-38, 
                         J.-P. Chen {\it et al.} (2011);
         K. Slifer, talk at the Tensor Polarized Solid Target Workshop.
\bibitem{eic} C. Weiss, N. Kalantarians, 
                  talks at the Tensor Polarized Solid Target Workshop.
\bibitem{azz} E. Long, M. Strikman, M. Sargsian, W. Cosyn,
                  talks at the Tensor Polarized Solid Target Workshop;
              T. Badman {\it et al.}, Proposal to Jefferson Lab PAC42.
\bibitem{fermilab-d} Xiaodong Jiang, 
           private communications on Fermilab Drell-Yan experiment (2014).
\bibitem{j-parc}
      See http://j-parc.jp/index-e.html for the J-PARC project.
      S. Kumano, Nucl. Phys. A {\bf 782}, 442 (2007); 
                 AIP Conf. Proc. {\bf 1056}, 444 (2008).
      Workshop on Hadron physics with high-momentum hadron beams 
      at J-PARC in 2013,  
                 http://www-conf.kek.jp/past/hadron1/j-parc-hm-2013/ .
\bibitem{gsi-fair} http://www.gsi.de/fair/index\_e.html.
\bibitem{feynman} R. P. Feynman,
                {\it Photon-Hadron Interactions} (Westview press, 1998). 
\bibitem{factor-2} The overall factor 1/2 is introduced in $b_1$
            as usual in defining $F_1$ and $g_1$ in terms of PDFs.
\bibitem{flavor3} S. Kumano, Phys. Rept. {\bf 303}, 183 (1998);
                  G. T. Garvey and J.-C. Peng,
                       Prog. Part. Nucl. Phys. {\bf 47}, 203 (2001);
	   J.-C. Peng and J.-W. Qiu, Prog. Part. Nucl. Phys. {\bf 76}, 43 (2014).
\bibitem{hkn07} M. Hirai, S. Kumano, and T.-H. Nagai,
                   Phys. Rev. C {\bf 76}, 065207 (2007)
                and references therein.
\bibitem{mstw08} A. D. Martin, W. J. Stirling, R. S. Thorne,
                 and G. Watt,  Eur. Phys. J. C {\bf 63}, 189 (2009).
                 The LO PDFs are used in this work.
\bibitem{rhic-d} E. D. Courant, report BNL-65606 (1998).
\end{thebibliography}
\end{document}